\documentclass{IEEEtran}
\usepackage{cite}
\usepackage{amsmath,amssymb,amsfonts}
\usepackage{graphicx}
\usepackage{textcomp,nicefrac}

\usepackage{siunitx} 
\usepackage{booktabs}
\usepackage{url}
\usepackage[caption=false, font=footnotesize, position=top]{subfig}

\newcommand{\dd}{\text{d}}
\newcommand{\inv}[1]{\frac{1}{#1}}

\newcommand{\paren}[1]{\left( #1 \right)}

\def\BibTeX{{\rm B\kern-.05em{\sc i\kern-.025em b}\kern-.08em
T\kern-.1667em\lower.7ex\hbox{E}\kern-.125emX}}
\begin{document}
\title{Performance of High-Intensity Electron Linacs as Drivers for Compact Neutron Sources}
\author{\IEEEauthorblockN{
Javier Olivares Herrador, Laurence M. Wroe, Andrea Latina, Roberto Corsini, Walter Wuensch, Steinar Stapnes, Nuria Fuster-Martínez, Benito Gimeno, Daniel Esperante}                                     
\thanks{J. Olivares Herrador, L. Wroe, A. Latina, R. Corsini and W. Wuensch are with the European Organization for Nuclear Research (CERN), Geneva, Switzerland. Email: javier.olivares.herrador@cern.ch.}
\thanks{J. Olivares-Herrador, N. Fuster-Martínez, B. Gimeno, D. Esperante are with Instituto de Física Corpuscular (IFIC-CSIC) - University of Valencia, Valencia, Spain.}
\thanks{D. Esperante is with the Department of Electronic Engineering, School of Engineering, University of Valencia, Spain.}
}
\maketitle

\begin{abstract}
The demand for neutron production facilities is increasing, as the applications of neutron science are manifold. These applications drive the need for efficient and compact neutron sources. In this context, this paper explores the potential performance of normal-conducting compact electron linacs from \qtyrange{20}{500}{MeV} as drivers for neutron sources. Results from a \textsc{G4beamline} simulation study find optimum dimensions of a tungsten target and characterize the emitted neutron spectrum. Two electron linac are then evaluated as drivers of such target: HPCI, an X-band linac; and CTF3 drive beam linac, an S-band alternative.
The simulation study shows neutron strength values up to $\qty{1.51 e15}{n/s}$ can be attained, as well as low energy consumed per neutron produced values up to \qty{5.65e-10}{J/n}, suggesting these electron-based neutron sources may be a compact and energy-efficient solution for many research, industry and medical applications.
\end{abstract}


\section{Introduction}
\label{sec:intro}
\par
\begin{figure}[!htb]
   \centering
    \resizebox{1.0\columnwidth}{!}{\input{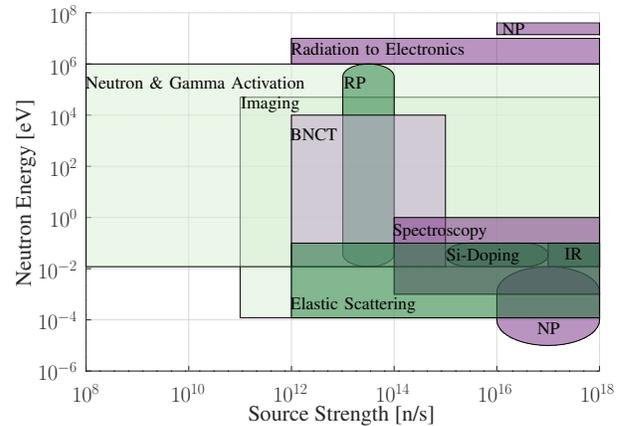}}
   \caption{Neutron applications chart. Figure from~\cite{my_ipac}. Data extracted from~\cite{neutrons_aapps} and~\cite{iaea}. The acronym BNCT stands for Boron Neutron Capture Therapy, NP for Nuclear Physics research, RP for Radioisotope Production, and IR for irradiation experiments.}
   \label{fig:neutron_application}
\end{figure}

 
Over the past century, nuclear reactors have been widely used neutron sources, with many laboratories and universities having benefited from them. Today, neutron sources continue to be utilized in a wide range of research topics and
applications, but plans have been made to shut down some of the existing reactors~\cite{nature}. As an alternative, accelerator-driven neutron sources are being designed and constructed to support and develop neutron activities~\cite{neutrons_aapps,iaea}. An overview of these applications and relevant  neutron energies and intensities is shown schematically  in Fig.~\ref{fig:neutron_application}.
\par 
\par 
Two types of accelerator-based neutron sources exist: spallation sources and Compact Accelerator-based Neutron Sources (CANS). The spallation process collides highly energetic protons (of hundreds of MeVs) against a high-$Z$ target typically made of lead, tungsten, or mercury. This results in the generation of large neutron fluxes ($\geq 10^{16}$ n/s), although at the expense of high power consumption (of some MWs)~\cite{Feizi_2016}.
\par 
CANS can be driven by either protons or electron accelerators. Proton-accelerator-based sources rely on a direct nuclear reaction of protons with a certain target, such as the $^7$Li(p,n)$^7$Be reaction~\cite{proton_reaction}. In the case of electron-accelerator-based sources, neutron production has already been demonstrated in facilities such as ORELA~\cite{orela} (which operated from 1969 to 2008) and GELINA~\cite{gelina} (currently in operation). These facilities respectively rely on  L-band (\qty{1.3}{GHz}) and S-band radiofrequency technologies (\qty{3.0}{GHz}) to accelerate beams up to \qty{150}{MeV}.
\par 
In the case of CANS driven by electron accelerators, the neutrons are produced via electron-induced spallation, where the electrons produce bremsstrahlung 
photons, which afterwards interact with the target nuclei to produce neutrons. For this reason, a high-Z material is required. Photonuclear neutron production can be summarized in three predominant mechanisms:
\begin{itemize}
    \item \textbf{Giant dipole resonance excitation:} Gamma photons from 7 to 40 MeV increase the dipole momentum of the nucleus as an ensemble. This extra energy leads to a large number of de-excitation events, which result in the emission of neutrons~\cite{gdr}. 
    \item \textbf{Quasi-deuteron disintegration:} Occurs when a photon of high energy (70-300 MeV) is absorbed by a pair of proton-neutron coming close together in the nucleus, forming a quasi-deuteron ensemble which disintegrates and emits a neutron~\cite{quasi-deuteron}. 
    \item \textbf{HEP effects:} Gamma photons above 150 MeV lead to the formation of energetic particles such as pions, which decay into high energy neutrons among other products~\cite{photo-pion}.
\end{itemize}
\par 
\par 
\par In this work, we explore the possibility of using modern high-intensity, normal-conducting, pulsed electron accelerators operating at full beam loading to produce neutrons in a compact and cost-effective facility. 
\par 
This paper is structured as follows: Section~\ref{sec:methods} presents the figures of merit for evaluating neutron production as well as the methodology for the analysis of neutron production, beam loading, and heat deposition. In Section~\ref{sec:tungsten}, the length and radius of a cylindrical tungsten target are optimized for maximum neutron production at different electron energies from \qtyrange{20}{500}{MeV}. 
In Section~\ref{sec:linacs}, two designs of normal-conducting electron accelerators are presented and reformulated to explore the challenges of high-intensity operation as drivers for neutron sources. Finally, Section~\ref{sec:state-of-art} discusses the performance of the proposed sources with respect to the state of the art.
\par 
\section{Methods}
\label{sec:methods}

\subsection{Neutron Production Figures of Merit}
\begin{figure}[h]
    \centering
    \includegraphics[width=0.7\columnwidth]{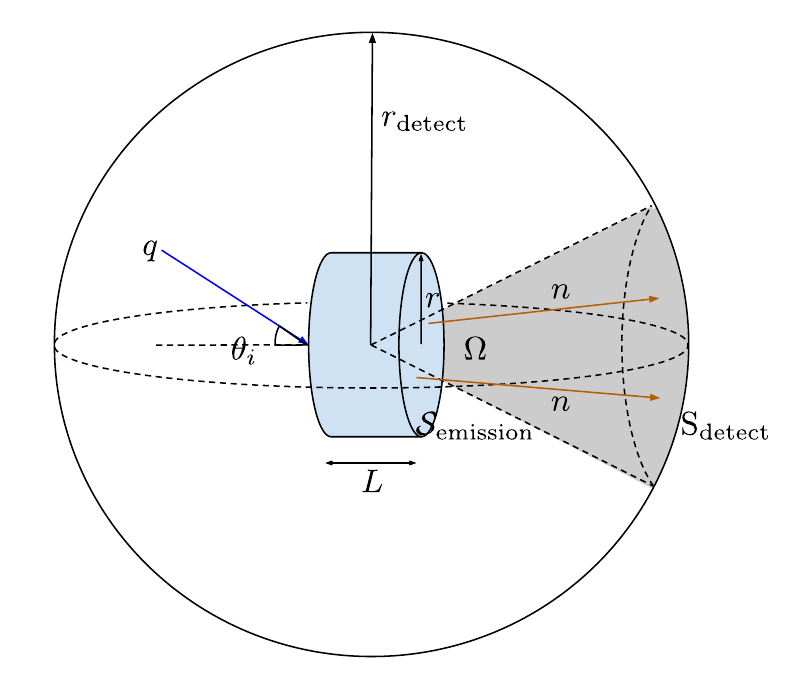}
    \caption{Spherical geometry for neutron production and detection when a charge $q$ hits the tungsten target (in blue).}
    \label{fig:tungsten_setup}
\end{figure}

To characterize neutron production, the following figures of merit are used in this paper:
\begin{itemize}
    \item The average neutron yield:        
    \begin{equation}
            Y_n(E_e) \equiv \frac{N_n(E_e)}{N_e} \ [\si{n/e}], \label{eq:yield}
        \end{equation}
        where $N_n$ is the total amount of produced neutrons for a given electron energy and $N_e$ the number of incident electrons.
        
    \item The solid-angle distribution, $f_\Omega$:
    \begin{gather}
        f_\Omega(\varphi, \theta_d; E_e) \equiv  \frac{\dd^2 Y_n}{\dd \Omega}(\varphi, \theta_d)  \ [\si{n/e/sr}] \label{eq:fomega},
    \end{gather}
    where $\dd \Omega(\theta_d,\varphi) = \sin{\theta_d} \dd \varphi \dd \theta_d$ is the solid angle of detection with $0 \leq \varphi < 2\pi$ and $0 \leq \theta_d < \pi$. Figure \ref{fig:tungsten_setup} shows a schematic of the neutron production setup, where the gray-area corresponds to the $\theta_d\in(0,30)^{\circ}$ area with $\varphi \in (0, 360)^{\circ}$. 
    \item The polar-angle distribution, $f_\theta$:
    \begin{equation}
        f_\theta(\theta_d; E_e) \equiv \inv{2\pi} \int_0^{2\pi} f_\Omega(\theta_d,\varphi) \dd \varphi \ [\si{n/e/sr}]. \label{eq:ftheta}
    \end{equation}
    \item The azimuthal energy spectrum of the emergent neutrons, $f_{\theta, E_n}$:
    \begin{equation}
        f_{\theta, E_n}(\theta_d, E_n; E_e) \equiv \frac{\text{d}f_\theta}{\text{d}E_n} \ [\si{n/e/sr/MeV}]. \label{eq:fE}
    \end{equation}
    \item The source strength as the average intensity of neutrons emitted in all directions
\begin{equation}
    I_n \equiv I_{e, \text{av}} Y_n \ [\si{n/s}],
    \label{eq:strength}
\end{equation} 
where $I_{e, \text{av}}$ is the average intensity of the electron accelerator. 
\end{itemize}

\par 
\subsection{\textsc{G4beamline} Simulations and Data Processing}
The interaction of the electron beam in tungsten is moddeled by \textsc{G4beamline}~\cite{g4beamline}, a software which uses Monte Carlo sampling to study radiation-matter interaction. 
The simulations undertaken in this study use the  physics option FTFP\_BERT\_HP, where hadron-nucleus interactions follow the FTF model~\cite{geant4-libraries}. In particular, the hadronic interaction of $<$~\qty{6}{GeV} gammas is handled by photonuclear processes which, among others, use the Bertini Cascade model~\cite{bertini} to take into account the effects mentioned in Sec. \ref{sec:intro}. The HP option activates the high-precision neutron models, which include an extensive set of libraries describing elastic and inelastic scattering, neutron capture and fission processes involving $<$~\qty{20}{MeV} neutrons~\cite{guerrero}.

\par
\color{black}
For each simulation setup, unless explicitly mentioned, the emergent neutrons were detected  on a sphere of radius $r_\text{detect}~=~$~\qty{1}{m} centred on the target. The neutron yield was computed as:
\begin{equation}
    Y_n = \inv{N_e}\sum_{i=1}^{N_e} \chi_i,
\end{equation}
where $\chi_i$ is the number of neutrons produced by the incident electron $i$. The uncertainty of the yield is calculated as $3\sigma_{Y_n}$, with~\cite{penelope}:
\begin{equation}
    \sigma_{Y_n} \equiv \sqrt{\frac{1}{N_e}Y_n(1-Y_n)}.
\end{equation}
\par 
The figures of merit in Eqs.\,(\ref{eq:fomega}-\ref{eq:ftheta}) refer to continuous distributions which are computed as histograms based on the associated stepwise constant functions. For instance, $f_\Omega$ was calculated as~\cite{penelope}:
\begin{equation}
    f_\Omega(\varphi,\theta_d;E_e) = f_{\Omega,j,k} \pm 3\sigma_{f_{\Omega,j,k}}
\end{equation}
for $\theta_{d,j-1} < \theta_d < \theta_{d,j}$ and $\varphi_{k-1} < \varphi < \varphi_k$. Here, $\theta_{d,j} = j\Delta \theta_d$, and $\varphi_k = k\Delta \varphi$, with $0<j,k<100$, $\Delta \theta_d = \frac{\pi}{99}$ rad and $\Delta \varphi = \frac{2\pi}{99}$ rad. Each bin value is expressed as:
\begin{align}
    f_{\Omega,j,k} = &  \frac{Y_{n,j,k}}{\sin{\theta_{d,j}}\Delta \theta_{d} \Delta \varphi} = \inv{N_e} \sum_{i=1}^{N_e} \frac{\chi_{i,j,k}}{\sin{\theta_{d,j}}\Delta \theta_{d} \Delta \varphi}, \label{eq:hist_fomega}\\
   \sigma_{f_{\Omega,j,k}} = & \inv{\sin{\theta_{d,j}}\Delta \theta_{d} \Delta \varphi}\sqrt{\inv{N_e}Y_{n,j,k}(1-Y_{n,j,k})}, \label{eq:hist_fomega_sigma}
\end{align}
where $\chi_{i,j,k}$ is the number of neutrons per electron produced by the $i$-th electron in the solid angle $\dd \Omega(\theta_{d,j},\varphi_k)$.
\par 
\subsection{Heat deposition calculations}
To calculate the deposited energy of the beam in the target, its volume was divided in $N_x=400$, $N_y=400$ and $N_z=10$ nodes, creating a mesh of $\{ (x_i,y_j,z_k) \}_{i,j}^{N_x,N_y}$ nodes where $x_i = i \frac{2r}{N_x-1} - r$, $y_j = j \frac{2r}{N_y-1} - r$, and $z_k = (k-0.5)L/N_z$. 
\par 
Denoting $e_{n,m,ijk}$ as the amount of energy deposited into the $(x_i, y_j, z_k)$ node by the $m-$th particle of the $n-$th shower, then the average energy deposited in the  $(x_i, y_j, z_k)$ node, $\Delta Q(i,j,k)$, and its uncertainty are obtained as~\cite{penelope}: 
\begin{align}
    \Delta Q(i,j,k) = & \inv{N_e}\sum_{n=1}^{N_e} e_{n,ijk}, \text{ with } e_{n,ijk} = \sum_m e_{n,m,ijk}, \\
    \sigma_{\Delta Q(i,j,k)} = & \sqrt{\inv{N_e} \paren{\inv{N_e}\sum_{n=1}^{N_e}e_{n,ijk}^2-\Delta Q(i,j,k)^2}}.
\end{align}

From this, the Peak value of the Energy Deposition Distribution (PEDD) in a pulse can be computed as:
\begin{align}
    Q_m(i,j,k) = & \frac{N_\text{bunches}q_\text{bunch}}{\rho e}\paren{\frac{\Delta Q(i,j,k)}{\Delta x \Delta y \Delta z}}, \\
    \text{PEDD} = & \max\{Q_m(i,j,k): 1 \leq i,j\leq N_x, 1\leq k \leq N_z \}. 
\end{align}
with $\Delta z = \frac{L}{N_z}$, $\Delta_x = \Delta_y =\frac{2r}{N_x-1} $, and $\rho$ the density of pure tungsten.
\subsection{\textsc{RF-Track} and Beam Loading Calculations}
\textsc{RF-Track}\cite{rf-track} is a tracking code developed at CERN which allows arbitrary particle tracking under the effect of external and self-induced forces. It includes a Beam Loading module~\cite{olivares} which  computes the accelerating gradient reduction in transient and steady regimes. 
\par 
\textsc{RF-Track} can be imported as an external library in \textsc{octave} and \textsc{python}. The \textsc{octave} optimization routines, used for BL compensation studies, are computed with the \textit{fminsearch} function using the Nelder-Mead simplex method~\cite{simplex}.
\section{Neutron Production from Electrons}
\label{sec:tungsten}
To optimize the dimensions of the tungsten target, the geometry and interaction illustrated in Fig.~\ref{fig:tungsten_setup} was simulated in \textsc{G4beamline} over a range of mean energies from \qtyrange{20}{500}{MeV}. The number of incident electrons simulated was  $10^8$ with the relative energy spread $\sigma_{E_e}/E_e$ set to $1\%$ and the incident angle, $\theta_i$, to $0^{\circ}$.
\par  
\subsection{Dimensions of the Tungsten Target}
\begin{figure}[t]
    \centering
    \includegraphics[width=0.8\linewidth]{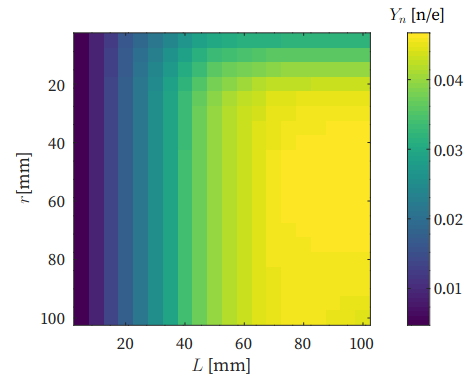}
    \caption{Dimension scan for the neutron yield of a cylindrical tungsten target when $\langle E_e \rangle$ = 500 MeV.}
    \label{fig:tungsten_optimization}
\end{figure}

Figure \ref{fig:tungsten_optimization} shows the neutron yield as a function of $L$ and $r$ for an incident electron beam of $\langle E_e \rangle = \qty{500}{MeV}$. An island of maximum production is visible for $L\geq $\qty{80}{mm} and $r \geq$ \qty{40}{mm} and  a similar island was observed at all electron energies. The optimal dimensions for maximum yield were therefore set at $L = 80$ mm and $r = 40$ mm. 
\par Table~\ref{tab:energy_results} presents the yield values with the optimal dimensions for mean incident electron energies $\langle E_e \rangle$ ranging from \qtyrange{20}{500}{MeV}. It shows that larger electron energies lead to greater neutron yields. This is consistent with the fact that the larger the electron energy is, the more intense the bremmstrahlung spectrum is~\cite{Feizi_2016}. This results in the creation of more photoneutrons. 
\par 
\begin{table}[t]
    \centering 
        \caption{$f_\theta$, and $\langle E_n \rangle$ for the neutrons emerging at $\theta_d = 130 ^{\circ}$ for different electron energies.}
        \begin{tabular}{cccc}
        \toprule
        $\boldsymbol{\langle E_e \rangle}$ &
         $\boldsymbol{Y_n}$ &
        $\boldsymbol{f_\theta}$ &  $\boldsymbol{f_{E_{n, \text{peak}}}}$  \\ 
        \textbf{[MeV]} & \textbf{[$ 10^{-1}$ n/e]} & \textbf{[$10^{-2}$  n/e/sr]} &  \textbf{[$10^{-2}$ n/e/sr/MeV]} \\ 
        \midrule
        20    & $0.0407 \pm 0.0021$  &  $0.045\pm 0.005$   &  $0.047 \pm 0.017$   \\
        50    & $0.31 \pm 0.05 $     &  $0.350 \pm 0.014$  & $0.33 \pm 0.04$   \\ 
        100   & $0.79 \pm 0.09 $    & $0.879 \pm 0.023$    & $0.77 \pm 0.7$   \\ 
        300    & $2.723 \pm 0.014  $ &  $2.89 \pm 0.04$    & $2.42 \pm 0.12$    \\ 
        500    & $4.644 \pm 0.015 $  &  $4.77 \pm 0.05$    & $3.89 \pm 0.15$    \\
        \bottomrule
    \end{tabular}
    \label{tab:energy_results}
\end{table}
\subsection{Angular distribution}
\begin{figure}[t]
    \centering
    \includegraphics[width=0.8\linewidth]{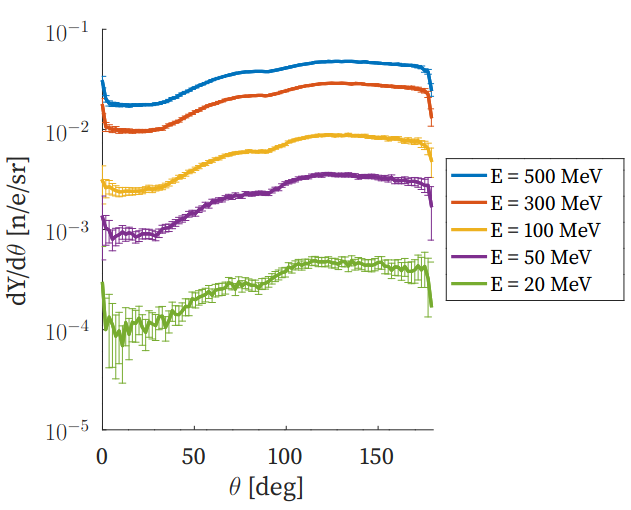}
    \caption{Neutron angular distribution for the optimal tungsten target ($L = 80$ mm, $r=40$ mm) for different $\langle E_e \rangle$ values.}
    \label{fig:tungsten_optimal_direction}
\end{figure}

Figure~\ref{fig:tungsten_optimal_direction} shows the polar angle distribution, $f_\theta$, as calculated for the optimal tungsten target for $\langle E_e \rangle$ values of $20$, $50$, $100$, $300$ and $500$ MeV. The results reveal that backward neutron production is larger than forward neutron production, with the direction of maximal production $\theta_d = 130^\circ$. Such direction exhibits a plateau of $\pm \qty{20}{^\circ}$, which allows stable neutron detection in this direction.
\par 
\par 
\subsection{Energy distribution}
\begin{figure}[t]
    \centering
    \includegraphics[width=0.8\linewidth]{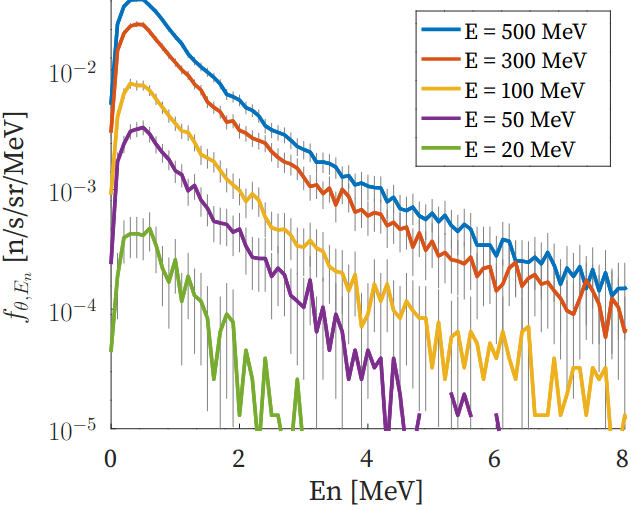}
    \caption{$f_{E_n,\theta}$ distribution for different $\langle E_e \rangle$ values at $\theta_d = 130^{\circ}$.}
    \label{fig:tungsten_EN_overview}
\end{figure}

The neutron energy spectrum produced by the optimized tungsten target is calculated according to Eq. (\ref{eq:fE}). Figure \ref{fig:tungsten_EN_overview} shows the results of $f_{\theta, E_n}$ for different $\langle E_e \rangle$ values at the direction of maximum emission, $\theta_d = 130 ^{\circ}$.
\par 
\par 
For all energies and all directions, the neutron spectrum is Maxwellian with a peak neutron energy around \qty{0.5}{MeV}, corresponding to the neutrons created mainly from the GDR. Larger electron energies, however, produce neutrons with larger $E_n$ values. The presence of a prominent peak in $f_{\theta, E_n}$ which does not vary with $\langle E_e \rangle$ translates into stability, since fluctuations in the electron beam energy do not alter such dominant value. 
\par 
From Fig. \ref{fig:neutron_application}, we see that these MeV neutrons could serve for electronics radiation studies, provided the source strength is large enough. For other applications including medical and industrial, these neutrons would require further moderation. 
\par 
\section{Full-beam-loaded electron linac sources}
\label{sec:linacs}
\par
Some neutron applications require pulsed spectra, and therefore benefit from pulsed linac operation. In this regard, CERN's ongoing efforts in developing normal-conducting linac technology, particularly within the framework of the Compact Linear Collider (CLIC)~\cite{clic}, provide a relevant basis for neutron production investigations. 

\par 
For this reason, we consider optimising the following linacs as potential drivers for CANS:
\par  
\begin{itemize}
\item The CLIC Test-Facility 3 (CTF3) drive-beam linac~\cite{ctf3}: A thermionic gun followed by bunching and accelerating S-band travelling-wave structures. It was operated between 2004 and 2016, and accelerated electrons up to \qty{150}{MeV}  to study the two-beam acceleration scheme of CLIC. 
\par 
\item The High-Pulse-Current-Injector (HPCI) linac~\cite{hpci}: A photoinjector followed by an X-band travelling-wave linac optimized to provide high-quality and high-current electron beams. The HPCI is a conceptual design based on CLIC X-band technology with potential applications in ICS sources or novel radiotherapy schemes \cite{carlo}.
\end{itemize}
Table~\ref{tab:electron-injectors} summarizes the relevant nominal parameters of the two high-intensity electron linacs.
\par 
\begin{table}[h]
\centering
\caption{Relevant nominal parameters of the linacs}
\begin{tabular}{lccc}
    \toprule
    \textbf{Magnitude}  & \textbf{Units}    & \textbf{HPCI linac~\cite{hpci}} & \textbf{CTF3 linac\cite{ctf3}} \\
    \midrule 
    $f$ & GHz & $12.00$ & $3.00$ \\ 
    $f_\text{RF-cycle}$& Hz   &    $100$  &  $100$  \\
    $r/Q$ & k$\Omega$/m (av) & $12.6$ & $4.40$ \\ 
    $Q$ & - (av) & $5717$ & $4000$ \\ 
    $v_g$ & $\%c$ (av) & $3.8$ & $3.5$\\ 
    $P_\text{max}$ & MW & $26.7$ & $29.8$\\ 
    \midrule 
    $q_\text{bunch}$    & nC    & $0.600$     & $2.33$ \\
    $N_\text{bunches}$  &       & $1000$    & $2100$  \\ 
    $ I_{e,\text{av}} $  &$\mu$A& $28.50$& $489.3$ \\ 
    \bottomrule
\end{tabular}
\label{tab:electron-injectors}
\end{table}
\par 
\subsection{Beam Loading}
Equation~(\ref{eq:strength}) shows that, for a given target at a given electron energy (i.e for a fixed yield value), the strength of a neutron source increases by transporting more charge through the electron linac. For a given input power, the maximum amount of charge that can be accelerated is limited mainly by beam loading (BL). 
As described in Ref.~\cite{olivares}, this effect scales with $\frac{\omega r/Q\tilde{I}}{2v_g}$, where $r/Q$ is the normalized shunt impedance per unit length, $v_g$ the group velocity, $\omega$ the resonating angular frequency, and $\tilde{I}$ the electron peak intensity. Comparing the frequency and $r/Q$ values in Table \ref{tab:electron-injectors}, it can be shown that, for the same bunch charge ($q_\text{bunch}$), beam loading has a more significant impact in the HPCI linac than in the CTF3 drive beam linac.
\par 
The maximum beam intensity is beam-loading limited at the point at which all EM energy is consumed and the structure cannot accelerate further bunches. This phenomenon is known as full beam loading. 
\par
\begin{figure}[h]
    \centering

    \begin{minipage}{0.49\textwidth}
        \begin{tabular}{cc}
        (a) & \\ &
        \includegraphics[width=0.8\linewidth]{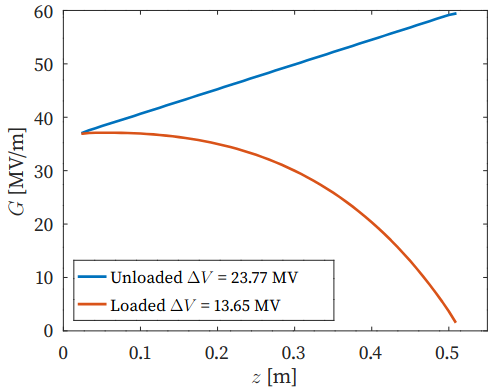} \\
        (b) & \\ &
        \includegraphics[width=0.8\linewidth]{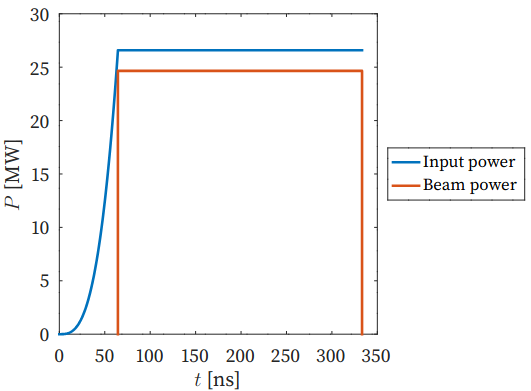} \\
\end{tabular}
\end{minipage}%
\caption{(a) Unloaded and steady-full-beam-loaded gradient of an HPCI X-band accelerating structure. (b) Optimized RF pulse and beam power for HCPI linac with $N_\text{bunches}$ = 1000, $q_\text{bunch}$ = \qty{602.4}{pC} and $f_b$ = \qty{3.00}{GHz.}}
\label{fig:gradient_comparison}
\end{figure}

 
CTF3 drive beam linac was designed to operate in a fully beam loaded mode, and its successful operation with $\geq 90\%$ percent of gradient reduction was proved in~\cite{ctf3-full-bl}. On the other hand, the  designed $q_\text{bunch}$ value of \qty{285}{pC} in Ref~\cite{hpci} for the HCPI linac does not reach the full beam loaded limit. For this reason, the BL-induced gradient reduction was simulated with \textsc{RF-Track} for the HPCI linac, and the full beam-loaded configurations for $95\%$ percent of gradient reduction were obtained for $\tilde{I}$ values of
\begin{equation}
    \tilde{I} = 1.8\,\si{A}.
\end{equation}
Since the HCPI injector is S-band, the proposed beam scheme is $q_\text{bunch} =$ \qty{600}{pC/bunch} with a bunch-to-bunch spacing of 100\,mm/$c$ (\qty{3}{GHz}). The laser and photocathode technology reported in~\cite{hpci} allows the consideration of trains of $1000$ bunches, large enough for the full BL steady response to manifest. Figure \ref{fig:gradient_comparison}a shows the gradient reduction at the steady state for the considered HPCI fully-loaded configuration, resulting in a loss of accelerating voltage of \qty{10.1}{MeV}.  
\par 
\subsection{BL compensation and RF-to-Beam Efficiency}
Without any dedicated compensation scheme, bunches in a train are accelerated by different amounts since the BL-induced gradient reduction takes place as the beam goes through. This induces a large beam energy spread, which could prevent the quadrupoles from providing the same focusing to all bunches, leading to beam instabilities which cause losses along the linac. For the CTF3 drive beam linac, BL compensation was achieved by appropriately delaying the arrival time of the rf pulse in the accelerating cavities with respect to the beam, as detailed in~\cite{ctf3-bl-compensation}. 
\par 
To perform BL compensation in the HPCI linac, Ref.~\cite{alexej} presents a compensation technique in TW structures based on a convenient modulation of input power that feeds the cavities, $P_\text{in}$. We used \textsc{RF-Track} and its Beam Loading module to replicate such compensation in simulation. Using a cubic polynomial to model $P_\text{in}$, we fixed its degrees of freedom to minimize the beam energy spread. The input power profile found is:
\begin{equation}
    P_\text{in}(t) = 
    \begin{cases} a_1 t^3 + a_2 t^2 + a_3 t \hspace{0.5cm} \text{for } t < t_\text{inj} \\ 
    P_\text{max} \hspace{2.43cm} \text{for } t \geq t_\text{inj} \label{eq:pinput}
    \end{cases}
\end{equation}
where $P_\text{max}$ is the maximum input power for which the structure is designed (see Table \ref{tab:electron-injectors}), $t_\text{inj}$ is the injection time of the beam in the structure with respect to the beginning of the RF pulse, and $a_1$, $a_2$, and $a_3$ are coefficients shown in Table~\ref{tab:BL_compensation_hpci}.
\par 
\begin{table}[h]
    \centering
    \caption{Beam Loading compensation parameters for HPCI linac.}
    \begin{tabular}{lcr}
    \toprule
         \textbf{Parameter} & \textbf{Units} & \textbf{Value} \\ 
    \midrule 
         $a_1$  &  \si{W/ns^3} & 98.3 \\
         $a_2$  & \si{W/ns^2}  & 41.4 \\ 
         $a_3$   & $\si{W/ns}$ &  261 \\ 
         $t_\text{inj}$ & ns & 63.5 \\
    \bottomrule
    \end{tabular}
    \label{tab:BL_compensation_hpci}
\end{table}

\par 
The input power profile, shown in Fig.~\ref{fig:gradient_comparison}b, allows the calculation of the instantaneous RF-to-beam efficiency, which is defined as:
\par 
\begin{equation}
    \eta(t) = \frac{\Delta P_\text{beam} (t)}{P_\text{in}(t)} = \frac{I(t)}{P_\text{in}(t)} \int_0^L G(z,t)\text{d}z. \label{eq:efficiency}
\end{equation}
\par 
In practice, many RF-to-beam efficiency estimations are based on the maximum value of $\eta$ achieved during a full pulse duration. However, To quantify RF-to-beam efficiency along the complete RF pulse, we define the average RF-to-Beam efficiency, $\eta_\text{av}$, as:
\begin{equation}
\eta_\text{av} = \frac{\int_0^{T_\text{pulse}} \Delta P_\text{beam}(t) \ \text{d} t}{\int_0^{T_\text{pulse}} P_\text{in}(t) \ \text{d}t} = \frac{P_\text{beam} (t_\text{out} - t_\text{in})}{\int_0^{T_\text{pulse}} P_\text{in}(t) \ \text{d}t}. \label{eq:av_efficiency}
\end{equation}
\par 
Table \ref{tab:rf-to-beam-efficiency} shows the RF-to-beam efficiency values for the heavy-loaded HPCI and CTF3 drive beam proposals.
\par 
\begin{table}[h]
    \centering
    \caption{RF-to-Beam efficiency values}
    \begin{tabular}{lcc}
        \toprule 
        \textbf{Facility} & $\boldsymbol{\max \eta}$ & $\boldsymbol{\eta_\text{av}}$ \\ \midrule 
        HPCI-linac & $0.93$ & $0.90$ \\
        CTF3 drive beam linac & $0.94$ & $0.91$ \\
        \bottomrule 
    \end{tabular}
    \label{tab:rf-to-beam-efficiency}
\end{table}
\par 

\par 

\subsection{Heat Deposition in the Target}
One of the potential issues associated with the production of neutrons with heavy-loaded electron linacs is the heat deposition in the target. Such high-intensity beams lead to a non-uniform energy deposition in the target. As a consequence, mechanical stresses may arise, expanding the target up to a threshold where the deformation is no longer elastic, causing target failures~\cite{positron_source_heat}. For pure tungsten, the PEDD threshold is \qty{35.1}{J/g}~\cite{clic}.
\par
\begin{figure}[h]
  \centering
\begin{minipage}{0.49\textwidth}
\begin{tabular}{cc}
(a) & (b) \\
   \includegraphics[width=0.45\linewidth]{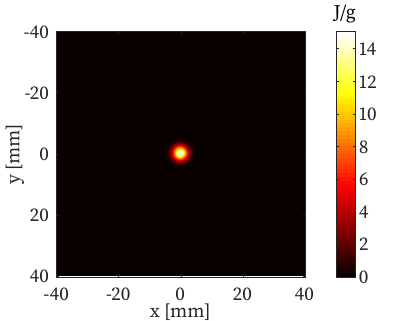} & 
    \includegraphics[width=0.45\linewidth]{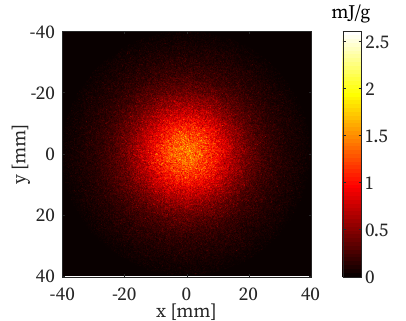} \\
\end{tabular}
\end{minipage}%
\hfill 
\begin{minipage}{0.49\textwidth}
\begin{tabular}{cc} 
     (c) & (d) \\
 \includegraphics[width=0.45\linewidth]{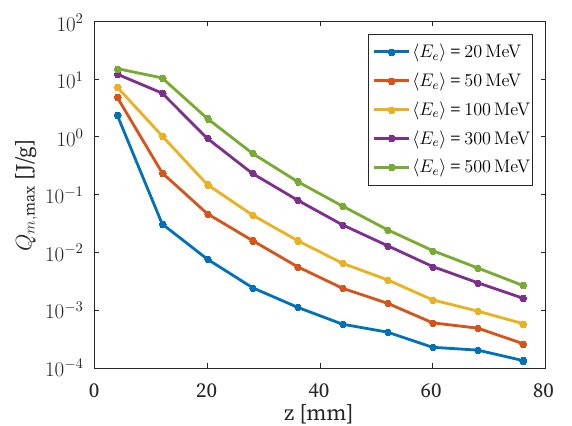} &
 \includegraphics[width=0.45\linewidth]{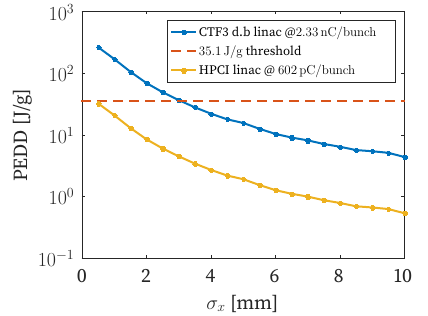} 
\end{tabular}
\end{minipage}
  \caption{Energy deposition in the tungsten target. (a) For HPCI \qty{600}{pC/bunch} beam with $\langle E_e \rangle =$ \qty{500}{MeV}, $\sigma_x = \sigma_y =$ \qty{1.3}{mm} at $z=\qty{0}{mm}$. (b) Idem at $z=\qty{80}{mm}$. (c) Energy deposition evolution with depth for HPCI beam configuration at $\sigma_x = \sigma_y = \qty{1.3}{mm}$. (d) Peak Energy deposition dependency on $\sigma_x$ for both e-linac-based sources proposed at $\langle E_e \rangle = \qty{500}{MeV}.$}
  \label{fig:heat_maps}
\end{figure}

 
To validate the HPCI- and the CTF3-drive-beam-linac-based CANS proposals, we examined the PEDD values with \textsc{G4beamline} simulations, using $N_e = 10^7$ events and following the procedure in Sec. \ref{sec:methods}. The results are shown in Figure~\ref{fig:heat_maps}.
\par 
\par 
Figure \ref{fig:heat_maps}c presents the maximum value of the deposited heat histogram in each detecting plane $z = z_k$ for the HPCI linac configuration (\qty{600}{pC/bunch}, $\sigma_x = \sigma_y = \qty{1.3}{mm}$). It reveals that the maximum heat deposition occurs at the beam entrance, which is due to the fact that electrons, as well as their products, spread out as they travel through the target, leading to a reduction in the concentration of heat deposition. 
In Fig. \ref{fig:heat_maps}d, the PEDD values for the two electron-linac-based neutron sources under consideration are depicted for different final beam sizes values, $\sigma_x$ at $\langle E_e \rangle = \qty{500}{MeV}$. The HPCI-linac-based proposal shows safe performance for $\sigma_x$ values larger than \qty{0.5}{mm}, while the CTF3-drive-beam option requires $\sigma_x \geq \qty{3.5}{mm}$.
\section{Comparison with the State of the art}
\label{sec:state-of-art}
So far, the proposed electron-linac-based neutron sources exhibit safe PEDD values, high RF-to-beam efficiency values and transport high $q_\text{bunch}$ values without reaching BL saturation. In this Section, we discuss how their performance compares with already-existing or under design sources in the state-of-the-art.
\par
\begin{figure}[htb]
    \centering
    \includegraphics[width=1\linewidth]{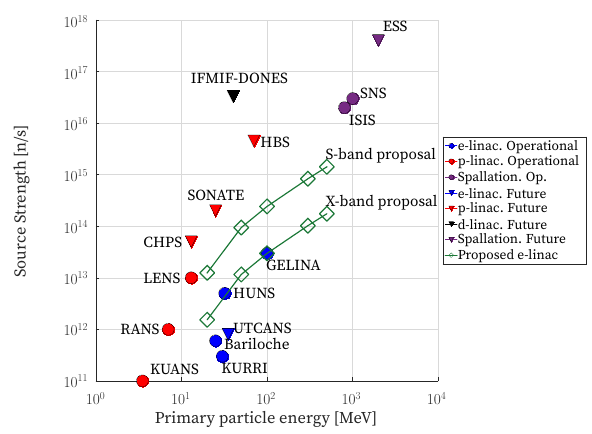}
    \caption{Source strength vs primary energy particle. Electron CANS (blue), proton CANS (red) data are extracted from Tab.\ref{tab:state-of-art-cans},  and spallation sources (purple) data are extracted from~\cite{ess},~\cite{sns},~\cite{isis}.  The uncertainty of the proposed e-linac CANS source strength (in green) is inferior to the marker size.}
    \label{fig:state_of_the_art1}
\end{figure}

 
Figure~\ref{fig:state_of_the_art1} is a landscape plot showing how the CTF3- and HPCI-driven CANS, with electron beams of $\langle E_e \rangle =$ $20,\, 50,\, 100,$ $300,\, 500$ MeV, compare to existing facilities in terms of source strength and primary particle energy.

\par 
\par 
The proposed electron-linac-based source's strength compares to some electron- and proton-linac-based compact sources, exhibiting a neutron strength up to $(1.51 \pm 0.07) \cdot 10^{15}$\,n/s. This limit in neutron production can be understood considering the power of the primary particle's beam, defined as:
\begin{equation}
      P_\text{beam} \equiv I_{e,\text{av}}\si{[A]} E_e\si{[eV]} / \text{e} \ \si{[W]}. \label{eq:power}
\end{equation}
\par
\begin{figure}[htb]
    \centering
    \includegraphics[width=1\linewidth]{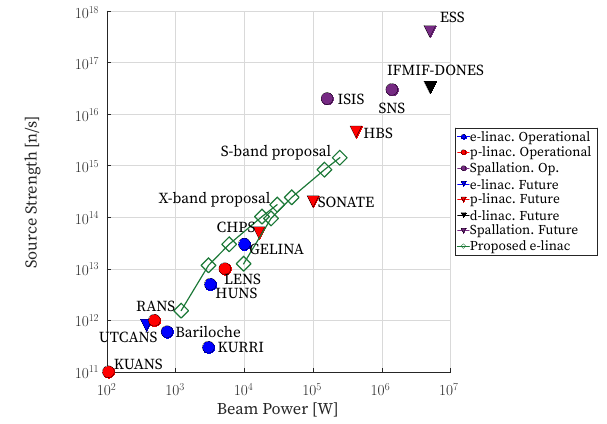}
    \caption{Strength vs average beam power for the proposed electron linacs and state-of-the-art sources. Electron CANS (blue), proton CANS (red) data are extracted from Tab.\ref{tab:state-of-art-cans},  and spallation sources (purple) data are extracted from~\cite{ess},~\cite{sns},~\cite{isis}. The uncertainty of the proposed e-linac CANS source strength (in green) is inferior to the marker size.}
    \label{fig:total_results_state_of_art_power}
\end{figure}

 
Figure~\ref{fig:total_results_state_of_art_power} shows the dependency of the source strength of the previous compact and spallation sources with the primary beam power. Among the considered sources in Fig.~\ref{fig:total_results_state_of_art_power}, the proposed e-linac facilities exhibit greater source strengths for beam powers ranging from $0.57$ kW to $245$ kW, meaning that the optimized tungsten target is more efficient than other targets in this power regime, regardless the nature of the particle arriving to it.
\par 
\par
\begin{figure}[htb]
    \centering
    \includegraphics[width=1\linewidth]{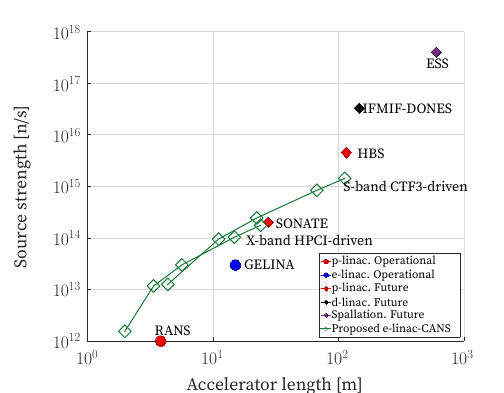}
    \caption{Source strength vs accelerator length for the proposed electron linacs and the state of art. The uncertainty of the proposed e-linac CANS source strength (in green) is inferior to the marker size. Data extracted from Tab.\ref{tab:efficiency} and \ref{tab:length_cost}.}
    \label{fig:state_of_art_length}
\end{figure}

 
Another quantity of interest to be considered when comparing different CANS is the accelerator length, $L_\text{acc}$. Figure ~\ref{fig:state_of_art_length} shows the dependency of the length of the accelerator-based sources with their strength. 
\par 
\par 
\subsection{Energy consumption and efficiency}
To estimate energy consumption in the neutron production process, the entire wall-to-beam process is considered. The HPCI linac and the CTF3 drive-beam linac share the same RF schematic: each klystron produces an RF pulse that is compressed with a pulse compressor, and then splits to feed two independent TW structures. Losses are expected in each of the steps, and the total wall-to-beam efficiency, $\eta_{_\text{TOTAL}}$ is calculated as the product of the efficiencies of each process. 
\par 
Table \ref{tab:efficiency} presents the total energy consumption for both electron-linac-based CANS proposals, as well as the energy consumed per neutron produced, $E_{\text{pn}}$. Both electron-linac-based sources exhibit similar wall-to-beam efficiencies, which translates into a similar energetic cost for neutron production, with X-band being more advantageous than S-band. 

\begin{table}[h!]

\caption{Efficiency considerations for $\langle E_e \rangle = \qty{500}{MeV}$.}
\begin{tabular}{lccc}
\toprule
\textbf{Magnitude} & \textbf{Unit} & \textbf{HPCI} & \textbf{CTF3 d.b} \\ 
\midrule
Bunch power gain $^{(1)}$ & MW & 24.6 & 26.6 \\
Pulse length & $\mu$s & 0.333 & 1.56 \\
Energy required $^{(2)}$ & J& 16.86 & 82.93  \\
\midrule
Loaded gradient & MV/m & 27.0 & 6.5 \\ 
Pairs of structures & & 16 & 28 \\
Total length & m & 24.0 & 112.1 \\
\midrule 
$\eta_{\text{splitter}}$  & - & 0.9 & 0.9 \\ 
$\eta_{\text{compressor}}$ & - & 0.56 \cite{carlo} & 0.66 \\ 
$\eta_{\text{wall-RF}}$ & - & 0.6 \cite{nuria} & 0.45 \\ 
\midrule
$\eta_{_\text{TOTAL}}$ & - & 0.27 & 0.24 \\ 
\midrule 
Train energy consumption & J & 986.2 & 9587 \\ 
RF Average power '' & kW & $98.62$ & $958.8$ \\ 
\midrule 
Yield & n/s & $1.74 \cdot 10^{14}$ & $1.51  \cdot 10^{15}$ \\
$E_\text{pn}$ & J/n & $5.65 \cdot 10^{-10}$ & $6.76  \cdot 10^{-10}$ \\
\bottomrule 
\end{tabular}
\label{tab:efficiency}
\begin{tabular}{cl}
    $^1$ & Per TW structure. \\
    $^2$ & From two TW structures. \\ 
    $^3$ & Estimation considering source and diagnostics. \\ 
\end{tabular}
\end{table}
\par 
\begin{figure}[t]
    \centering
    \includegraphics[width=1\linewidth]{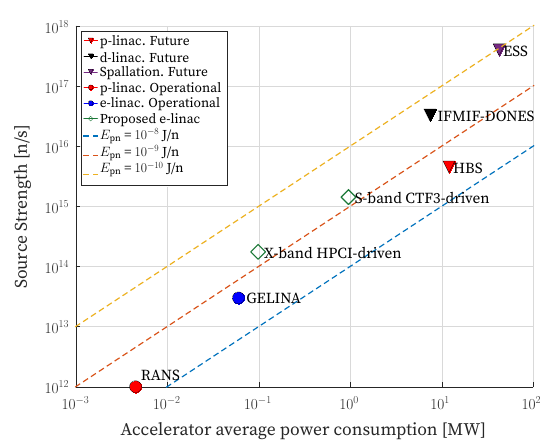}
    \caption{Source strength vs RF average power consumption for the proposed e-linacs at $\langle E_e \rangle = \qty{500}{MeV}$ and the state of the art facilities in Tab. \ref{tab:length_cost}. }
    \label{fig:energy_efficiency}
\end{figure}
Figure \ref{fig:energy_efficiency} compares the average RF power consumption of different accelerator-based neutron sources with respect to their source strength. It shows that the proposed electron-linac-based sources exhibit $E_\text{pn}$ values one order of magnitude smaller than the others compact sources. 
\par 
\begin{table}[h]
\centering
\caption{State-of-the-art: Length and power consumption.}
\begin{tabular}{lcc}
    \toprule 
    \textbf{Facility} & $\boldsymbol{L_\text{acc}}$\textbf{ [m]} & \textbf{RF Average power} \\ 
    & & \textbf{consumption [MW]} \\ 
    
    \midrule 
    RANS & $3.8$~\cite{rans_length} & $4.5\cdot 10^{-3}$~\cite{rans_power} \\ 
    GELINA~\cite{linacs_recap} & $15.0$ & $6.0\cdot 10^{-2}$ \\ 
    HBS~\cite{hbs} & $115.4\,^{(1)}$ & $12.0$ \\ 
    IFMIF DONES & $141.6$~\cite{ifmif_design} & $7.4$ $^{(2)}$~\cite{ifmif_rf}\\ 
    ESS~\cite{ess_tech_report} & $602.4$& $42.1$ $^{(2)}$ \\
    \bottomrule 
\end{tabular}
\\ 
\footnotesize{
\begin{tabular}{cl}
    $^1$ & Assumed \qty{0.5}{m} drift between structures. \\
    $^2$ & Cryogenic cooling system power consumption included. \\ 
\end{tabular}
}
\normalsize
\label{tab:length_cost}
\end{table}

\footnotesize{
\begin{table}[b!]
 \caption{State-of-the-art compact accelerator-driven neutron sources}
\centering
 \begin{tabular}{ccc}
\hline 
\textbf{Source} & \textbf{Driver} & \textbf{Source strength }              \\ 
               &                 & $\boldsymbol{I_n}$ , $\boldsymbol{E_\text{range}}$   \\ \hline
%
GELINA  & S-band e-linac + W & $3.4 \cdot 10^{13}$ \\ 
Belgium~\cite{gelina} & $100$\,MeV / $\SI{100}{\micro A}$ & $10^{-4} - 10^6$ eV \\
\hline 
HUNS              & S-band e-linac + Pb            & $5 \cdot 10^{12}$ n/s    \\
Japan~\cite{jcans} & $32$\,MeV / $\SI{100}{\micro A}$ &      $1 - 10^{-4}$   eV                \\
\hline 
KURRI-LINAC      & L-band e-linac + Ta  & $3 \cdot 10^{11}$  \\
Japan~\cite{jcans}    & $46$\,MeV / $\SI{1}{\micro A}$  &                \\
\hline 
Bariloche Linac & e-Linac + Pb & $6 \cdot 10^{11}$ n/s     \\ 
(Argentina)~\cite{springer} & $25$\,MeV / $\SI{30}{\micro A}$ &                    \\ 
\hline
RANS               & p-linac + Be          & $10^{12}$ n/s           \\
Japan~\cite{jcans} & $7$\,MeV / $\SI{70}{\micro A}$ & $10^3 - 10^{-4}$ eV                                 \\
\hline 
KUANS               & p-linac + Be            & $10^{11}$ n/s                            \\
Japan~\cite{jcans}  & $3.5$\,MeV / $\SI{30}{\micro A}$ & $10^3 - 10^{-2}$ eV                           \\ 
\hline 
LENS                     & p-linac + 2RFQ + Be & $10^{13}$ n/s      \\
USA~\cite{springer} & $13$\,MeV/ $20$ mA   &                   \\
\hline 
\hline 
\multicolumn{3}{c}{\textbf{Sources under study or ongoing development}} \\
\hline 
\hline
CPHS & p-Linac + Be           & $5\cdot 10^{13}$ n/s  \\
China~\cite{cphs} & 16\,MeV / $\SI{1.25}{\micro A}$  &            \\
\hline 
LENOS  & p-RFQ + Li       &  $10^{15}$ n/s  \\
Italy~\cite{springer}  & $5$\,MeV/ $50$ mA &  not moderated  \\ 
\hline  
IFMIF-DONES                & d-linac+RFQ+SRF+Li          & $5\cdot 10^{14}$ n/s/cm$^2$  \\ 
Spain~\cite{ifmif}   & $40$\,MeV / $125$ mA  & @source, forward  \\
\hline
UTCANS        &  X-band e-linac + W        & $8 \cdot 10^{11}$ n/s  \\
Japan\cite{springer}  & $35$\,MeV / $250$ mA (peak) & (C$_2$H$_4$)$_n$ moderator \\
\hline  
SONATE & p-linac+Li/Be & $3.1 \cdot 10^{13}$ n/s\\
France~\cite{sonate} & $20$\,MeV/ $100$ mA &  \\
\hline
HBS                & p-linac+ Ta           & $9.1 \cdot 10^{14}$ n/s/mA \\
Germany~\cite{hbs} & $70$\,MeV / $100$ mA & $\langle E_n \rangle = 0.45$\,MeV       \\
\hline 
\end{tabular}
\label{tab:state-of-art-cans}
\end{table}
} \normalsize

\section{Conclusions}
This paper shows that normal-conducting high-intensity electron linacs are suitable drivers for compact neutron sources, which offer a compact, affordable, and competitive alternative compared to existing medium-flux fission or spallation-based sources. In particular, we presented the optimization and characterization of a tungsten target based on extensive \textsc{G4beamline} simulations, and examined the performance of two the state-of-the-art normal conducting S-band and X-band electron linear accelerators. 
\color{black}
\par 
Maximizing the bunch charge up to the full beam loading regime allows maximizing the neutron emission as well as the RF-to-beam efficiency. However, despite the high intensity, safe values of heat deposition in the target are found provided that the focusing is not too strong ($\sigma_x \geq \qty{3.5}{mm}$, well below the target radius). 
\par 
The comparison with the state of the art shows that electron-accelerator-based sources are, in general, equally compact and more efficient than proton-based or deuteron-based sources, making them valuable alternatives for moderate neutron production scenarios, like medical and industrial applications.  
\par 
In particular, without further moderation scheme, the proposed electron-linac-based sources offer very energetic neutron beams of the order of the MeVs as seen in Table~\ref{tab:energy_results}. Therefore, according to Fig.~\ref{fig:neutron_application}, the discussed facilities could serve mostly for electronic irradiation studies such as~\cite{matteo_clear}. 
\par 
Most industrial and medical applications benefit from less energetic neutrons, thus requiring moderation. Therefore, further work should aim at the design of a target-moderator-assembly for a particular application. 
\bibliographystyle{IEEEtran}
%
\bibliography{refs2} 

\begin{thebibliography}{10}
\providecommand{\url}[1]{#1}
\csname url@samestyle\endcsname
\providecommand{\newblock}{\relax}
\providecommand{\bibinfo}[2]{#2}
\providecommand{\BIBentrySTDinterwordspacing}{\spaceskip=0pt\relax}
\providecommand{\BIBentryALTinterwordstretchfactor}{4}
\providecommand{\BIBentryALTinterwordspacing}{\spaceskip=\fontdimen2\font plus
\BIBentryALTinterwordstretchfactor\fontdimen3\font minus
  \fontdimen4\font\relax}
\providecommand{\BIBforeignlanguage}[2]{{%
\expandafter\ifx\csname l@#1\endcsname\relax
\typeout{** WARNING: IEEEtran.bst: No hyphenation pattern has been}%
\typeout{** loaded for the language `#1'. Using the pattern for}%
\typeout{** the default language instead.}%
\else
\language=\csname l@#1\endcsname
\fi
#2}}
\providecommand{\BIBdecl}{\relax}
\BIBdecl

\bibitem{nature}
J.~M. Carpenter, ``The development of compact neutron sources,'' \emph{Nature
  Reviews Physics}, vol.~1, no.~3, pp. 177--179, 2019.

\bibitem{neutrons_aapps}
Y.~Kiyanagi, ``Neutron applications developing at compact accelerator-driven
  neutron sources,'' \emph{AAPPS Bulletin}, vol.~31, pp. 1--19, 2021.

\bibitem{iaea}
\BIBentryALTinterwordspacing
\emph{Compact Accelerator Based Neutron Sources}, ser. TECDOC Series.\hskip 1em
  plus 0.5em minus 0.4em\relax Vienna: INTERNATIONAL ATOMIC ENERGY AGENCY,
  2021, no. 1981. [Online]. Available:
  \url{https://www.iaea.org/publications/14948/compact-accelerator-based-neutron-sources}
\BIBentrySTDinterwordspacing

\bibitem{my_ipac}
J.~O. Herrador, L.~Wroe, A.~Latina, W.~Wuensch, R.~Corsini, S.~Stapnes,
  N.~Fuster-Martinez, B.~Gimeno, and D.~Esperante, ``Neutron production using
  compact linear accelerators,'' in \emph{15th International Particle
  Accelerator Conference Proceedings}, 2024, pp. 673--676.

\bibitem{Feizi_2016}
\BIBentryALTinterwordspacing
H.~Feizi and A.~Ranjbar, ``Developing an accelerator driven system (ads) based
  on electron accelerators and heavy water,'' \emph{Journal of
  Instrumentation}, vol.~11, no.~02, p. P02004, feb 2016. [Online]. Available:
  \url{https://dx.doi.org/10.1088/1748-0221/11/02/P02004}
\BIBentrySTDinterwordspacing

\bibitem{proton_reaction}
\BIBentryALTinterwordspacing
M.~S. Herrera, G.~A. Moreno, and A.~J. Kreiner, ``Revisiting the 7li(p,n)7be
  reaction near threshold,'' \emph{Applied Radiation and Isotopes}, vol.~88,
  pp. 243--246, 2014, 15th International Congress on Neutron Capture Therapy
  Impact of a new radiotherapy against cancer. [Online]. Available:
  \url{https://www.sciencedirect.com/science/article/pii/S0969804313004880}
\BIBentrySTDinterwordspacing

\bibitem{orela}
T.~S. Bigelow, C.~Ausmus, D.~Brashear, K.~Guber, J.~Harvey, P.~Koehler, R.~B.
  Overton, J.~A. White, and V.~Cauley, ``Recent operation of the orela electron
  linac at ornl for neutron crosssection research,'' in \emph{Proceedings of
  the Linear Accelerator Conference (LINAC)}, no.~21, 2006, pp. 79--81.

\bibitem{gelina}
\BIBentryALTinterwordspacing
D.~Ene, C.~Borcea, S.~Kopecky, W.~Mondelaers, A.~Negret, and A.~Plompen,
  ``Global characterisation of the gelina facility for high-resolution neutron
  time-of-flight measurements by monte carlo simulations,'' \emph{Nuclear
  Instruments and Methods in Physics Research Section A: Accelerators,
  Spectrometers, Detectors and Associated Equipment}, vol. 618, no.~1, pp.
  54--68, 2010. [Online]. Available:
  \url{https://www.sciencedirect.com/science/article/pii/S0168900210005589}
\BIBentrySTDinterwordspacing

\bibitem{gdr}
F.~H. Lewis and J.~D. Walecka, ``Electromagnetic structure of the giant dipole
  resonance,'' \emph{Phys. Rev.}, vol. 133, pp. B849--B868, Feb 1964.

\bibitem{quasi-deuteron}
S.~Fujii, ``Note on the quasi-deuteron model for nuclear
  photo-disintegration,'' \emph{Il Nuovo Cimento (1955-1965)}, vol.~25, pp.
  995--1007, 1962.

\bibitem{photo-pion}
V.~Petwal, V.~Senecha, K.~Subbaiah, H.~Soni, and S.~Kotaiah, ``Optimization
  studies of photo-neutron production in high-z metallic targets using high
  energy electron beam for ads and transmutation,'' \emph{Pramana}, vol.~68,
  pp. 235--241, 2007.

\bibitem{g4beamline}
T.~Roberts, ``G4beamline user’s guide,'' \emph{Muons, Inc}, pp. 3468--3470,
  2013.

\bibitem{geant4-libraries}
\BIBentryALTinterwordspacing
C.~web repository, ``Geant4 libraries official website, consulted on the 7th of
  october 2024.'' [Online]. Available:
  \url{https://geant4.web.cern.ch/documentation/dev/plg_html/PhysicsListGuide/reference_PL/FTFP_BERT.html}
\BIBentrySTDinterwordspacing

\bibitem{bertini}
D.~Wright and M.~Kelsey, ``The geant4 bertini cascade,'' \emph{Nuclear
  Instruments and Methods in Physics Research Section A: Accelerators,
  Spectrometers, Detectors and Associated Equipment}, vol. 804, pp. 175--188,
  2015.

\bibitem{guerrero}
E.~Mendoza, D.~Cano-Ott, T.~Koi, and C.~Guerrero, ``New standard evaluated
  neutron cross section libraries for the geant4 code and first verification,''
  \emph{IEEE Transactions on Nuclear Science}, vol.~61, no.~4, pp. 2357--2364,
  2014.

\bibitem{penelope}
F.~Salvat, J.~M. Fern{\'a}ndez-Varea, J.~Sempau \emph{et~al.}, ``Penelope-2006:
  A code system for monte carlo simulation of electron and photon transport,''
  in \emph{Workshop proceedings}, vol.~4, no. 6222.\hskip 1em plus 0.5em minus
  0.4em\relax Citeseer, 2006, p.~7.

\bibitem{rf-track}
A.~Latina, ``Rf-track reference manual,'' CERN, Geneva, Switzerland, Tech.
  Rep., 2024.

\bibitem{olivares}
\BIBentryALTinterwordspacing
J.~Olivares~Herrador, A.~Latina, A.~Aksoy, N.~Fuster~Martínez, B.~Gimeno, and
  D.~Esperante, ``Implementation of the beam-loading effect in the tracking
  code rf-track based on a power-diffusive model,'' \emph{Frontiers in
  Physics}, vol.~12, 2024. [Online]. Available:
  \url{https://www.frontiersin.org/journals/physics/articles/10.3389/fphy.2024.1348042}
\BIBentrySTDinterwordspacing

\bibitem{simplex}
J.~A. Nelder and R.~Mead, ``A simplex method for function minimization,''
  \emph{The Computer Journal}, vol.~7, no.~4, pp. 308--313, 01 1965.

\bibitem{clic}
M.~Aicheler, P.~Burrows, M.~Draper, T.~Garvey, P.~Lebrun, K.~Peach, N.~Phinney,
  H.~Schmickler, D.~Schulte, and N.~Toge, ``A multi-tev linear collider based
  on clic technology: Clic conceptual design report,'' SLAC National
  Accelerator Lab., Menlo Park, CA (United States), Tech. Rep., 2014.

\bibitem{ctf3}
G.~Geschonke and A.~Ghigo, ``Ctf3 design report,'' Tech. Rep., 2002.

\bibitem{hpci}
A.~Latina, V.~Muşat, R.~Corsini, L.~A. Dyks, E.~Granados, A.~Grudiev,
  S.~Stapnes, P.~Wang, W.~Wuensch, E.~Cormier, and G.~Santarelli, ``A compact
  inverse compton scattering source based on x-band technology and
  cavity-enhanced high average power ultrafast lasers,'' in \emph{67th ICFA
  Adv. Beam Dyn. Workshop Future Light Sources Conference Proceedings}, 2023,
  pp. 257--260.

\bibitem{carlo}
\BIBentryALTinterwordspacing
C.~R. et~al., ``\BIBforeignlanguage{English}{The deep electron flash therapy
  facility},'' in \emph{\BIBforeignlanguage{English}{Proc. 32nd Linear
  Accelerator Conference (LINAC2024)}}, ser. International Linear Accelerator
  Conference, no.~32.\hskip 1em plus 0.5em minus 0.4em\relax JACoW Publishing,
  Geneva, Switzerland, 08 2024, paper WEYA002, pp. 551--556. [Online].
  Available: \url{https://indico.jacow.org/event/71/contributions/5203}
\BIBentrySTDinterwordspacing

\bibitem{ctf3-full-bl}
\BIBentryALTinterwordspacing
R.~Corsini, M.~Bernard, G.~Bienvenu, H.~Braun, G.~Carron, A.~Ferrari,
  O.~Forstner, T.~Garvey, G.~Geschonke, L.~Groening, E.~Jensen, R.~Koontz,
  T.~Lefèvre, R.~Miller, L.~Rinolfi, R.~Roux, R.~D. Ruth, D.~Schulte, F.~A.
  Tecker, L.~Thorndahl, and A.~D. Yeremian, ``{First Full Beam Loading
  Operation with the CTF3 Linac},'' 2004. [Online]. Available:
  \url{https://cds.cern.ch/record/791372}
\BIBentrySTDinterwordspacing

\bibitem{ctf3-bl-compensation}
A.~Dabrowski, S.~Bettoni, H.~Braun, E.~Bravin, R.~Corsini, S.~Doebert,
  C.~Dutriat, T.~Lef{\`e}vre, M.~Olveg{\aa}rd, P.~Skowronski \emph{et~al.},
  ``Transient beam loading compensation in ctf3,'' in \emph{LINAC08, Victoria,
  BC, Canada}, 2008, pp. 585--587.

\bibitem{alexej}
\BIBentryALTinterwordspacing
A.~Lunin, V.~Yakovlev, and A.~Grudiev, ``{Analytical solutions for transient
  and steady state beam loading in arbitrary traveling wave accelerating
  structures},'' \emph{Phys. Rev. Spec. Top. Accel. Beams}, vol.~14, no.~5, p.
  052001, 2011. [Online]. Available: \url{https://cds.cern.ch/record/1333709}
\BIBentrySTDinterwordspacing

\bibitem{positron_source_heat}
\BIBentryALTinterwordspacing
X.~Artru, V.~N. Baier, R.~Chehab, M.~Chevallier, M.~S. Dubrovin, A.~Jejcic, and
  J.~Silva, ``{Positron Sources Using Channeling: A Comparison with
  conventional Targets},'' \emph{Part. Accel.}, vol.~59, pp. 19--41, 1998.
  [Online]. Available: \url{http://cds.cern.ch/record/1120307}
\BIBentrySTDinterwordspacing

\bibitem{ess}
\BIBentryALTinterwordspacing
``Ess official website,'' accessed the 6th of November, 2023. [Online].
  Available: \url{https://europeanspallationsource.se/}
\BIBentrySTDinterwordspacing

\bibitem{sns}
\BIBentryALTinterwordspacing
``Sns official site,'' accessed the 10th of May, 2024. [Online]. Available:
  \url{https://neutrons.ornl.gov/sns}
\BIBentrySTDinterwordspacing

\bibitem{isis}
\BIBentryALTinterwordspacing
``Isis official website,'' accessed the 9th of November, 2023. [Online].
  Available: \url{https://www.isis.stfc.ac.uk/Pages/home.aspx}
\BIBentrySTDinterwordspacing

\bibitem{nuria}
P.~Alonso~Arias, A.~Chauchet, C.~Marrelli, I.~Syratchev, M.~Webber, M.~Boronat,
  M.~Jones, N.~Catalan-Lasheras, S.~Gonz\'alez-Ant\'on, and U.~N. Zaib,
  ``{Validation of high efficiency klystron technology},'' \emph{JACoW}, vol.
  LINAC2024, p. THPB015, 2024.

\bibitem{rans_length}
\BIBentryALTinterwordspacing
D.~L. Friesel and W.~Hunt, ``Performance of an accsys technology pl-7 linac as
  an injector for the iucf cooler injector synchrotron,'' in \emph{Proceedings
  of 19th International Linear Accelerator Conference (LINAC’98), Chicago,
  IL, US}, 1998. [Online]. Available:
  \url{https://api.semanticscholar.org/CorpusID:54781112}
\BIBentrySTDinterwordspacing

\bibitem{rans_power}
Y.~Otake, ``Riken accelerator-driven compact neutron systems, rans project and
  their capabilities,'' \emph{Neutron News}, vol.~31, no. 2-4, pp. 32--36,
  2020.

\bibitem{linacs_recap}
J.~Clendenin, L.~Rinolfi, K.~Takata, and D.~Warner, ``Compendium of scientific
  linacs,'' \emph{CERN/PS}, vol.~96, p.~32, 1996.

\bibitem{hbs}
T.~Br{\"u}ckel, T.~Gutberlet, J.~Baggemann, S.~B{\"o}hm, P.~Doege, J.~Fenske,
  M.~Feygenson, A.~Glavic, O.~Holderer, S.~Jaksch \emph{et~al.},
  \emph{Conceptual Design Report-J{\"u}lich High Brilliance Neutron Source
  (HBS)}.\hskip 1em plus 0.5em minus 0.4em\relax Forschungszentrum J{\"u}lich
  GmbH, Zentralbibliothek, Verlag, 2020.

\bibitem{ifmif_design}
W.~Kr{\'o}las, A.~Ibarra, F.~Arbeiter, F.~Arranz, D.~Bernardi, M.~Cappelli,
  J.~Castellanos, T.~D{\'e}zsi, H.~Dzitko, P.~Favuzza \emph{et~al.}, ``The
  ifmif-dones fusion oriented neutron source: evolution of the design,''
  \emph{Nuclear Fusion}, vol.~61, no.~12, p. 125002, 2021.

\bibitem{ifmif_rf}
\BIBentryALTinterwordspacing
D.~Regidor, C.~{de la Morena}, D.~Iriarte, F.~Sierra, S.~Dragaš, P.~Marini,
  J.~Sanz, J.~Molla, and A.~Ibarra, ``Ifmif-dones rf system,'' \emph{Fusion
  Engineering and Design}, vol. 167, p. 112322, 2021. [Online]. Available:
  \url{https://www.sciencedirect.com/science/article/pii/S0920379621000983}
\BIBentrySTDinterwordspacing

\bibitem{ess_tech_report}
K.~Andersen, ``Ess technical design report,'' 2012.

\bibitem{jcans}
\BIBentryALTinterwordspacing
``Jcans web site,'' accessed the 9th of November, 2023. [Online]. Available:
  \url{https://www.jcans.net/rans.html}
\BIBentrySTDinterwordspacing

\bibitem{springer}
I.~Anderson, C.~Andreani, J.~Carpenter, G.~Festa, G.~Gorini, C.-K. Loong, and
  R.~Senesi, ``Research opportunities with compact accelerator-driven neutron
  sources,'' \emph{Physics Reports}, vol. 654, pp. 1--58, 2016.

\bibitem{cphs}
T.~Sano, J.-i. Hori, Y.~Takahashi, H.~Yashima, J.~Lee, and H.~Harada,
  ``Analysis of energy resolution in the kurri-linac pulsed neutron facility,''
  in \emph{EPJ Web of Conferences}, vol. 146.\hskip 1em plus 0.5em minus
  0.4em\relax EDP sciences, 2017, p. 03031.

\bibitem{ifmif}
I.~Podadera, A.~Ibarra, D.~Jim{\'e}nez-Rey, J.~Molla, C.~de~la Morena,
  C.~Oliver, N.~Bazin, N.~Chauvin, J.~Dumas, S.~Chel \emph{et~al.}, ``The
  accelerator system of ifmif-dones multi-mw facility,'' in \emph{Proceedings
  of 12th International Particle Accelerator Conference (IPAC’21), Campinas,
  Brazil}, 2021, pp. 1910--1913.

\bibitem{sonate}
F.~Ott, A.~Menelle, and C.~Alba-Simionesco, ``The sonate project, a french cans
  for materials sciences research,'' in \emph{EPJ Web of Conferences}, vol.
  231.\hskip 1em plus 0.5em minus 0.4em\relax EDP Sciences, 2020, p. 01004.

\bibitem{matteo_clear}
G.~Lerner, A.~Coronetti, J.~M. Kempf, R.~G. Alía, F.~Cerutti, D.~Prelipcean,
  M.~Cecchetto, A.~Gilardi, W.~Farabolini, and R.~Corsini, ``Analysis of the
  photoneutron field near the thz dump of the clear accelerator at cern with
  seu measurements and simulations,'' \emph{IEEE Transactions on Nuclear
  Science}, vol.~69, no.~7, pp. 1541--1548, 2022.

\end{thebibliography}

\end{document}